\documentstyle{article}

\font\tenrm=cmr10
\font\tenit=cmti10
\font\elevenbf=cmbx10 scaled\magstep 1
\font\elevenrm=cmr10 scaled\magstep 1
 1

\renewenvironment{thebibliography}[1]
 { \elevenrm
   \begin{list}{\arabic{enumi}.}
    {\usecounter{enumi} \setlength{\parsep}{0pt}
     \setlength{\itemsep}{3pt} \settowidth{\labelwidth}{#1.}
     \sloppy
    }}{\end{list}}

\begin{document}
\begin{flushright} UH-511-782-94 \\ February 1994
\end{flushright}
\begin{center}
\vglue 0.6cm
 {\elevenbf        \vglue 10pt
\vglue 3pt
{Solar and Atmospheric Neutrinos \\ }  
\vglue 5pt
\vglue 1.0cm
{\tenrm Sandip Pakvasa \\}
\baselineskip=13pt
{\tenit Department of Physics and Astronomy, University of
Hawaii at Manoa \\}
\baselineskip=12pt
{\tenit Honolulu, HI 96822, USA\\}}
\end{center}

\baselineskip=14pt
\elevenrm


\begin{center}
\vspace{6mm}
\noindent{\normalsize{\underline{ABSTRACT}}}
\end{center}

Possible explanations of solar neutrino
and atmospheric neutrino
anomalies are summarized and future tests discussed.

\section{\bf Introduction}

\qquad In the standard model (SM) with no singlet right-handed
$\nu's$ and a single Higgs field, all neutrino masses are
zero and lepton number (as well as individual flavor quantum
numbers) are exactly conserved.  It follows that the charged
leptonic current is diagonal in both mass and flavor basis
and the mixing angles are zero.  Hence any evidence for
non-zero neutrino masses or for non-trivial mixings is
evidence for physics beyond the Standard Model.  This makes
the search for neutrino masses and mixings doubly important:
as measurement of fundamental parameters of intrinsic
interest and as harbingers of new physics.  In this talk I
will concentrate on hints from solar and atmospheric
neutrino observations that suggest non-zero neutrino masses
and mixings.

\section{\bf{{Solar Neutrinos}}}

\qquad The current status of the data on solar neutrino
observations from the four on-going experiments is
summarized in the table 1.

\begin{center}
Table I \\
The solar neutrino data
\cite{Suzuki,Anselmann,Abazov,Davis}
compared to the SSM\cite{Bahcall1}
\begin{tabular}{ll}
\\ \hline \hline
Experiment      &    Data/SSM \\ \hline
Kamiokande      &    $0.51 \pm 0.07$ \\
Gallex          &    $0.66 \pm 0.12$  \\
Sage            &    $0.44 \pm ^{0.17}_{0.21}$ \\
Homestake       &    $0.28 \pm 0.04$
\\ \hline \hline
\end{tabular}
\end{center}

The Kamiokande detector is
sensitive only to $^8B$ neutrinos; and the Homestake detector
is sensitive to $^8B \ (77\%)$ as well as $^7Be (14\%)$, pep
(2\%) and CNO (6\%) neutrinos \cite{Bahcall1}.  If the observations need no
new neutrino properties, then the $^8B$ $\nu's$ are not
distorted in their spectrum and the flux seen by Kamiokande
(over a limited energy range), can be assumed uniform and
hence applicable to Homestate as well.  In that case a
minimum of $(38 \pm 8)\%$ of SSM counting rate is contributed
by $^8B$ neutrinos alone and adding pep neutrinos it is
$(40\pm 8)$\% to be compared to the observed $(28 \pm 4)$\%.
It is obvious that something must reduce the $^7Be$
neutrino flux drastically to obtain agreement.  Since the
effective temperature dependence of $^7Be \ \nu$ flux is much
weaker than for $^8B$ flux \cite{Bahcall2}, it is difficult to arrange for a
stronger suppression for $^7Be$ than for the $^8B$ flux.  This
is borne out in calculations where the core temperature is
allowed to be a free parameter and it is found that a good
fit to all the data cannot be obtained \cite{Bludman}.  Furthermore, no
solar model has been found which can reproduce the Chlorine
rates even with the reduced $^8B$ flux or even come close
\cite{Bahcall3}.
There is a general agreement that with the Chlorine data
averaged over the whole period some neutrino properties are
called for \cite{Bahcall4}.

Even if the remaining uncertainties in the solar modelling
(or very low energy nuclear cross-sections)
and difficulties inherent in pioneering experiments may
cloud the interpretation of solar neutrino data in terms of
neutrino properties\cite{Morrison};
it is important to keep in mind that
there is no question that neutrinos from the sun have been
detected: both at high energies - 10 MeV (Kamiokande,
Homestake) and at
low energies - 1 MeV (Gallex, SAGE).  Hence a powerful
neutrino beam with sensitivity to $\delta m^2 \geq 10^{-10}
eV^2$ and $\sin^2 2 \theta \geq 0.1$ is available, free of
charge. It behooves us to utilise this beam maximally; and
future upcoming experiments will do just that.  They have
rates of order $10^4$ per year; in real time, spectrum
measurement, flux monitoring (via NC/CC in SNO) and low
threshold (in Borexino).  If the neutrino parameters lie in
this region we will definitely know the answer by 1996.
\newpage
\begin{center}
Table II \\
Future Detector Characteristics
\end{center}
\begin{tabular}{lllllll}
Status  & Detector  &  Size & $E_{th}$  &  Ev/yr  &  Reaction
                                        & Features \\
   &          &    &   &  &   &         \\

        &           &    &    & \small 10,000   & \small $\nu_e D$
          $\rightarrow$ \small epp   &    \small spectrum \\
\normalsize Constr. &  \small  SNO(\^{c})   &  \small 1KT  &
\small 5 MeV  &  \small 3000 & \small $\nu~D$ \small
$\rightarrow$
\small $\nu$\small pn  &   \small NC/CC   \\
         &          &      &    & \small 500     &   \small $\nu e$ &     \\

\normalsize Constr. &  \small SuperK(\^{c})  & \small 22KT  &
\small 5 MeV  &  \small 8000   &  \small $\nu e$
                                         &  \small spectrum  \\
        &                  &       &     &   \small      &     \\
\normalsize Test    &   \small Borexino (LS) & \small 0.2KT   &
\small 0.25 \ \small MeV     & \small 15,000
&  \small $\nu e$    & \small $^7Be$ \ \small Line  \\
\normalsize Test    &   \small ICARUS (IC)   &  \small 3KT  &
\small 5 \ \small MeV   &  \small 2700    &
\small $\nu e$     \\
	&    &         &  &  \small 3000     &  \small $\nu_e Ar
\rightarrow$ $eK^*$  &   \small spectrum  \\
        &   &         &     &      &   \quad \quad \quad \ \

\small  $\rightarrow \gamma$ &  \ \small \quad \\

\small  Prop.  &   \small Hellaz    &   \small  12T   &
\small
0.1 MeV    &  \small 12000   &  $\nu e$     & \small pp  \\
&   &   &  &  &  &  \\
&  &  &  &  &  &
\end{tabular}

Assuming that neutrino properties are the culprit,
I will summarize the solutions to the solar neutrino deficit
with emphasis on the non-MSW
options.  For definiteness and
simplicity I will assume (i) SSM fluxes of Bahcall
and Pinsounnent, (ii) two flavor mixing,
(iii) and ignore mixing with sterile neutrinos and
neutrino flavor changing neutral currents.  I will briefly discuss
the solutions and how each may be distinguished in future
experiments; especially in Borexino, SNO, Superkamiokande
and ICARUS \cite{These}.

{\bf MSW:}

This is the case in which $\delta m^2$ and $\sin^22 \theta$
lie in the range in which the solar matter effects are very
important \cite{Wolfenstein}.  A fit to all four experiments leaves three
allowed regions \cite{Anselmann2}.  One is the small angle $(\sin^22 \theta
\sim 4.10 ^{-3}, \delta m^2 \sim 10^{-5} eV^2)$ region; in
this region the rate for $^7Be \ \nu e$ scattering in Borexino
varies rapidly between $0.2$ and $0.5$ of SSM
and $^8B$ spectrum as
seen in SNO or Superkamiokande will show distortion.
Another is the large angle large $\delta m^2$ region $(\sin^22
\theta \sim 1, \delta m^2 \stackrel{\sim}{>} 10^{-5} eV^2)$;
in this region $^7Be$ is suppressed between 0.35 and 0.7 and
there is no distortion of $^8B$ spectrum.  Finally there is
a small region at large angle small $\delta m^2 (\sin^2 2
\theta \sim 1, \delta m^2 \stackrel{\sim}{<} 10^{-6} eV^2)$;
here there is a strong day-night variation in $^7Be$ line as
seen in Borexino \cite{Raghavan}.

{\bf Large Angle Long Wavelength:}

   The large angle long wavelength ("just so") \cite{Acker1}
continues to fit
all the data \cite{Acker2} with $\delta m^2 \sim 10^{-10} eV^2$ and
$\sin^2 2 \theta \stackrel{\sim}{>}0.8$.  Matter effects are
negligible.  This has striking predictions testable with
future detectors:
(i) suppression of $^7Be$ in Borexino between 0.2 and
0.5,
(ii)sharp distortion of $^8B$ spectrum and most
importantly,
(iii) visible oscillations of $^7Be$ line with time
of the year with upto factor of 2 variations.  This maybe
the only chance \cite{Pakvasa1} to see true quantum mechanical neutrino
oscillations [Fig. 1] and can be easily seen in Borexino and
distinguished from the $1/r^2$ variation.

Akhmedov et al. \cite{Akhmedov} have given an interesting possible
justification of such a scenario.  They suppose that
(i) there are only LH $\nu's$,
(ii) lepton number is conserved except by gravity;
then at Planck scale there may be lepton number violating
terms such as

\begin{equation}
\frac{g_{ij}}{m_{p}} \ \bar{\psi} {^{c \tau}_{Li}}_{-}
\psi_{Lj} \cdot \ \bar{\phi} ^{\tau}_{-} \phi
\end{equation}
where $\phi$ is the standard Higgs doublet, $m_p$ Planck
mass, i and j are family indices.  Then the neutrino masses
are Majorana and the mass matrix is
\begin{equation}
M_{\nu_{ij}} = g_{ij} v^2/m_p
\end{equation}
If one makes the further assumption that gravity is
flavor-blind and $g_{ij} = g$ and $g \sim 0(1)$ then the
matrix is
\begin{eqnarray}
m_\nu = \frac{v^2}{m_p}
\left (
\begin{array}{ccc}
1 & 1 & 1 \\
1 & 1 & 1 \\
1 & 1 & 1 \end{array} \right )
\end{eqnarray}
which has as mass eigenvalues
$m_1 = 0, m_2 = 0$ and $m_3 = m= 2 v^2/m_p  \cong 10
^{-5} eV$.  Hence $\delta m^2$ is about $10^{-10} eV^2$.
The mixing matrix is easily calculated and it can be shown
that
\begin{equation}
P(\nu_e \rightarrow \nu_e, L) \ = 1 - \frac{8}{9} \sin^2
\frac{m^2L}{4E}
\end{equation}
which corresponds to an effective $\sin^2 2 \theta$ of 0.89.

{\bf Decay with Mixing:}

\ A very old proposal is to have the neutrinos decay on the
way to the earth \cite{Bahcall5}.  The SN1987A observation of
$\bar{\nu_e}'s$ require that there be a stable component
in $\nu_e$ and the mixing be not too small \cite{Frieman}.  There must be
also some new physics for the decay into another neutrino
and a light or massless boson.  In any case,
phenomenologically, with the most recent Kamiokande and
Gallex
data in hand, the decay solution is ruled out at 98\%
C.L.\cite{Acker3}.

{\bf Matter induced Decay:}

There is another kind of decay that has been discussed in
the literature; this is the matter induced or MSW-
catalyzed decay\cite{Berezhiani}.  The basic idea is that in matter the
effective mass for $\nu_e$ is
greater than for $\bar{\nu}_e$ and if a coupling to a scalar (e.g.
Majoron) $\chi$ existed than the decay $\nu_e \rightarrow
\bar{\nu}_e + \chi$ could occur in matter (but not in
vacuum).  Similarly, in presence of a flavor changing
coupling the more relevant decay
$\nu_e \rightarrow \bar{\nu}_\mu + \chi$ can also occur in
matter.  This matter induced decay lifetime (Lab) behaves as
a constant rather than $E_\nu$ and hence this remains a
viable solution.

\newpage

{\bf Flavor Violating Gravity:}

If gravitational interaction of neutrinos is not diagonal in
flavor then even for massless neutrinos there are
oscillations induced by this flavor dependent gravitational
potential \cite{Gasperini}.  The survival probability for $\nu_e$ is
given by
\begin{equation}
P(\nu_e \rightarrow \nu_e, L) \ = 1- sin^2 \ (2 \theta_G)
\sin^2 [\delta \bar{\phi} EL]
\end{equation}
where $\bar{\phi}$ is the gravitational potential averaged
over the neutrino path-length, $\delta$ is the departure from
flavor independence of gravity:
$\delta = f_e - f_\mu$.

The quantity $(\delta \phi E L)$ can be written as $(\pi L/
\lambda _{G})$ where  \linebreak
$\lambda _{G} =6 km (\frac{10^{-20}}{\delta
\phi}) [\frac{1}{E/10 GeV}].$
The precise value of $\phi$ at the earth and the sun is very
uncertain due to potentially large contributions from
"nearby" large masses such as the Virgo cluster or the
local super cluster.  Current limits on $\delta \phi$ from
re-interpretations of $\delta m^2 - sin^2 2 \theta$ bound
are (for $\nu_e - \nu_x) \ 10^{-19}$ for $\sin^2 2 \theta_G
\sim 1.$  It turns out that $\delta \phi$ in the range
$10^{-20} - 10^{-21}$ and $\sin^2 2 \theta_G \sim 1$ can
provide a simultaneous good fit to all the solar neutrino
data as well as the atmospheric anomaly.  Future
long-baseline experiments can extend the bounds on $\delta
\phi$ to $10^{-22}$ or better and test this hypothesis
as proposed by Pantaleone et al \cite{Gasperini}.

To summarize, future detectors such as SNO,
Superkamiokande, Borexino, and ICARUS will have real time event
rates of several thousand per year.  They will measure the
$^8B$ neutrino energy spectrum accurately, $^7Be$ line rate
and the ratio of NC/CC in $\nu_e D$ reaction.  With this
information at hand it should be possible to establish that (a)
neutrino properties are relevant, (b) distinguish between
MSW, long wavelength, decay etc.,  (c) pin down the
parameters narrowly and (d) deduce more precise information
about the sun such as the core temperature.

{\section{\bf {Atmospheric
Neutrinos}}}

\quad Neutrinos are produced by cosmic rays interacting in the
atmosphere.  A primary (P) reacts with ``air" nucleus as:

\begin{equation}
``P" \ + \mbox{``air"}  \rightarrow \pi + x
\end{equation}
The $\pi$ may interact or decay; if it decays:

\begin{equation}
\pi \rightarrow \mu + \nu_\mu
\end{equation}
and at low energies (few GeV) the $\mu$ can also decay
before it hits the ground:
\begin{equation}
\mu \rightarrow e + \nu_e + \nu_\mu
\end{equation}
If all the $\mu's$ decay we are led to expect $N
(\nu_\mu)/N(\nu_e)$ of 2 (ignoring the distinction between
$\nu$ and $\bar{\nu})$.  This ratio, furthermore, is
expected to be essentially independent of the zenith angle
at low energies.  Neutrinos of energies below 2 GeV give
rise to ``contained" events in typical kiloton underground
detectors.  The results from the two large water-Cerenkov
detectors suggest that the ratio $R= N(\nu_\mu)/N(\nu_e)$ is
smaller than expected by almost a factor of two.  Kamiokande
finds (based on 6.1 Kton yr) for the ratio of
ratios\cite{Hirata}.

\begin{equation}
R_{obs}/R_{MC} = 0.60 \pm 0.07
\end{equation}
while IMB finds (based on 7.7 Kton yr)\cite{Casper}

\begin{equation}
R_{obs}/R_{MC} = 0.54 \pm 0.07
\end{equation}

\noindent The result of Frejus (for contained events) and Nusex is
respectively $0.87 \pm 0.21$ (based on 1.56 Kton yr)
and $0.99 
 \pm 0.40 ($ based on 0.4 Kton yr)
\cite{Berger,Aglietta}.
Finally, very recently SOUDAN II has found a result
of $0.69 \pm 0.19$ based on a 1 Kton yr- exposure
\cite{Litchfield}.

The ratio $N(\nu_\mu) /N(\nu_e)$ is considered more reliably
calculated than the individual fluxes: the ratio is stable
to about 5\% amongst different calculations whereas the
absolute fluxes vary by as much as 20 to 30\%
\cite{Volkova}.

The most important question is whether there is a ``mundane"
explanation for the deficit or is a new physics explanation
called for?  Let us consider the mundane explanations. (i)
Perhaps the $e/\mu$ identification in the water Cerenkov
detectors is simply wrong.  In response to this, Kamiokande
has made a very convincing case for the correctness of their
$e/\mu$ identification by showing how it works very well in
finding the expected number of $\mu \rightarrow e$ decays in
their contained events \cite{Suzuki1}.  Also, the fact that Soudan-II sees
the same deficit (in a non-water-Cerenkov detector) is
encouraging.  Finally, the upcoming (underway?) beam test at
KEK should settle the issue once and for all. (ii) There is
the question of low energy $\nu$-nuclear cross-sections and
lepton energy distributions in the region $E_\nu \sim$ 200
MeV to 1 GeV.  Ideally we would like to have these
$(\nu^{16} 0 \rightarrow \ell ^{16}F)$ measured
experimentally.  Even though $e/\mu$ universality is not
expected to be violated except kinematically (and hence in a
known manner) the difference between $\nu$ and $\bar{\nu}$
cross sections is important.  This is unlikely to be the
explanation\cite{See}.
(iii) It has been pointed out by
Volkova\cite{Volkova1} that if $\pi^+$ at low energies
dominates over $\pi^-$ then (because
$\sigma_{\bar{\nu}_{e}} < \sigma_{\nu_{e}})$ the effect is to
enhance $e/\mu$ signal.  She finds that with a
$\pi^+/\pi^- \sim 2.5$ (compared to values in the range
1.1-1.3 used to by others) the effect is only about 10\% of
the observed.  The importance
of knowing the relative amount of $\bar{\nu}_e/\nu_e$ has
also been stressed by Suzuki\cite{Suzuki1}.

Now we turn to new physics solutions for the anomaly.  The
first two will depend on assuming some absolute flux
calculation which accouns for the $\nu_\mu$ flux correctly;
e.g. the flux calculated by Bugaev and Naumov\cite{Bugaev}.
(i) The simplest explanation\cite{Learned} is that there is a universal
$\nu_e$ excess of flux $10^{-3} cm^{-2} GeV^{-2} sr^{-1}
sec^{-1}$.  Its spectrum may be falling like, say, $E^{-2}$
or $E^{-3}$; in that case the energy density is about 1/100
of that in cosmic rays and is quite ``safe".  These must be
isotropic (according to observations) and could be galactic
or more likely extra-galactic.  Could they be from AGN's?
(ii) A very interesting proposal
by Mann et al.\cite{Mann} is that the excess "e" events are not due to
$\nu_e's$ at all but due to proton decay mode $P \rightarrow
e^+ \nu \nu$.  In this case, the energy spectrum of the
excess e events should end at $E_e \sim 600
\ MeV$.  Both Kamiokande and IMB have a few events beyond 600
MeV but the errors are large and the hypothesis can not be
ruled out with the present data.  The rate corresponds to a
$\tau/BR$ of $4.10^{31}$ years.

To account for such a mode, the theoretical model has to
forbid other decay modes (in addition to predicting this
mode).  The simplest operator for such a decay mode (in
absence of a light $\nu_R)$ is $f/M 6 < \phi >uud \
{\overline{e^+ \nu \nu}}$ and hence $M \sim 10^5$ GeV.  In some typical
models this implies Leptoquarks in mass range below 1 TeV
and discoverable at LHC or Linear colliders\cite{Rudaz}.

A very recent calculation of the atmospheric neutrino flux
by Perkins\cite{Perkins}uses new muon measurements in the atmosphere
by MASS collaborations\cite{Perkins}.  He finds that the absolute
fluxes tend to agree with Barr et al.\cite{Volkova} but do not support
the Bugaev-Naumov\cite{Bugaev} low fluxes needed for the
interpretations given above.  For absolute fluxes somewhere
between Bugaev-Naumov and Barr et al., another
interpretation is possible viz. a universal isotropic source
of equal number of $\nu_e's$ and $\nu'_\mu s$, as suggested
e.g. by Tomozawa\cite{Tomozawa}.

We now turn to the flux independent explanation in terms of
neutrino oscillations.
The deviation of $R_{obs}/R_{MC}$ from 1 is fairly uniform
over zenith
angle and is most pronounced in the charged lepton energy
range 200-700 MeV which corresponds to neutrino energies
from 300 MeV to 1.2 GeV.  If we are to interpret this
deficit of $\nu_\mu's$ (and/or excess of $\nu_e's)$ as being
due to neutrino oscillations, the relevant parameters are
determined rather easily \cite{Learned1}.
The typical height of production,
h, is about 15-20 km above ground and for a zenith angle
$\theta$ the distance travelled by the neutrino before
reaching the detector is
\begin{equation}
L(\theta) = R \left [ {\sqrt{(1 + h/R)^2 - \sin^2 \theta}} -
     \cos \theta \right ]
\end{equation}
where $R$ is the radius of the earth.
Allowing for angular smearing due to the scattering and
finite angular resolution one finds that neutrino path
lengths can vary between 30km to 6500 km, and hence $L/E$
can vary between 25 km/GeV and 20,000 km/GeV.  Since the
data do not show any L (i.e. $\theta)$ or $E$ dependence we
may infer that the oscillations have already set in at
$E_\nu \sim 1 \ GeV$ and $L \sim 30 km$ and hence $\delta m^2$
cannot be much smaller than $3.10^{-2}eV^2$.  As for the
mixing angle $\theta$, if $P$ denotes the average
oscillation probability i.e. $P = \sin^2 2 \theta
< \sin^2 \delta m^2 L/4E > \approx \frac{1}{2} \sin^2 2
\theta$; then $R \ = 1-P$ in case of $\nu_\mu - \nu_\tau$
oscillations and for $\nu_\mu - \nu_e$ oscillations
\begin{equation}
R= \frac{1-(1-r)P}{1+(1/r -1)P}
\end{equation}
where $r = N(\nu_e)/N(\nu_\mu)$ in absence of oscillations
and most flux calculations yield $r \sim 0.45$.  Since $R$
is nearly 0.6, large  mixing angles of order $30^0$ to
$45^0$ are called for, $\nu_\mu - \nu_e$ mixing
needing somewhat smaller ones.
Detailed fits by Kamiokande and more recently IMB,
bear these expectations out
although somewhat bigger range of parameters $(\delta m^2$ up to
$4.10^{-3} eV^2$ and mixing angles up to $20^0$)
are allowed \cite{Hirata}.

There are also higher energy muons in the underground
detectors.  Typically in IMB and Kamiokande detectors events
are classified as thrugoing muons and stopping muons.  The
average $\nu_\mu$ energy for these correspond to about 100
GeV and 10 GeV respectively.  These events are expected to
have the famous $sec \theta$ zenith angle distribution due
to the competition between $\pi$ decay and interaction and
the $\nu_e$ flux is very small since the high energy $\mu's$
do not have time to decay in 20 km \cite{Gaisser}.
If the above explanation
of the low energy anomaly is correct then for the thrugoing
events (a) the zenith angle distribution should be distorted
since for horizontal events oscillations will not have set
in $(\delta m^2 L/4E \ll 1)$ but for vertical events there
should be depletion (b) the total muon event rate itself
should be decreased by the depletion and (c) in case of
$\nu_\mu-\nu_e$ oscillations there should be an enhancement
of $\nu_e$ (and hence showering) events especially at
energies where there might be matter
enhancement\cite{Learned1,Acker4}. Several
detectors, IMB, Kamiokande, Baksan, KGF (and now MACRO)
have data of the order
of a few hundred events
each\cite{Becker,Oyama,Boliev,Duggal,Barish}.
There is no clear distortion
of the zenith angle distribution or depletion of the
total rate seen in any data.  However, since the
comparison has to be made to absolute flux calculations, the
limits on $\delta m^2, \theta$ derived are not yet strong
enough to rule out the values needed to explain the low energy
anomaly \cite{Frati}.  IMB has derived
forbidden regions \cite{Becker} by taking ratio of
stoppers/thrugoers which is largely flux
independent and which rules
out the large angle region $(\sin ^2 2 \theta \sim 0.6$ to 1)
for $\delta m^2 \sim 3.10^{-3}$ to $8.10^{-3}$.  The same
data when used to constrain $\nu_\mu - \nu_e$ mixing
yield very weak constraints\cite{McGrath}.  In any case, neutrino
oscillation explanation of the contained event anomaly is
not ruled out.

\section{\bf Simultaneous Explanations for Solar and Atmospheric
Anomalies:}


\quad \ \ \ (i) If
$\nu_\mu - \nu_\tau$ mixing is responsible for atmospheric
deficit with $\delta m^2 \sim 10^{-2} eV^2$ and large
mixing;  and if the see-saw mechanism\cite{Yanagida} is operative, with
$m{_{\nu_\tau}} \sim 0.1 \ eV$ one expects a $m_{\nu_{\mu}}$
in the range $( \frac{mc}{mt})^2 m_{\nu_{\tau}} \sim
3.10^{-3} eV$ and leads to $\delta m^2$ for
$\nu_\mu-\nu_e$ of about $10^{-5} \ eV^2$,
making it just right for an explanation of solar neutrino
data via MSW.  In this case neutrino masses are all less
than 0.1 eV and there is no possibility to account for any
hot dark matter without sterile neutrinos.

(ii) If the atmospheric anomaly is accounted by
$\nu_\mu-\nu_e$ mixing with $\delta m^2 \stackrel{\sim}{<}
\ 0(10^{-2})eV^2$ and the see-saw mechanism is operative; then
the $\nu_\tau$ mass is of order of $(m_t/m_c)^2
m_{\nu_{\mu}} \sim 10 \ eV$.  This is in the right range to
account for hot dark matter in the mixed (30\% HDM, 70\% CDM)
DM scenario to account for the large scale structure which
has been proposed recently\cite{Klypin}.  The implications for solar
neutrinos are a uniform energy dependent depletion by about
$(1- \frac{1}{2} \sin^22 \theta_{e_{\mu}}) \sim 0.5$ to
0.75.  This is completely consistent with Kamiokande and
Gallium results but not with $^{37}Cl$ results.  In any case
it can be verified in the future solar neutrino detectors.
Both of the above scenarios were first discussed by Learned
et al. in 1988\cite{Learned1}.

(iii) {\bf Three flavor mixing:}

A very interesting possibility is that $\delta
m^2_{21} \sim 10^{10} eV^2$ and $\delta m^2_{31} \sim
\delta m^2_{32} \sim 10^{-2}eV^2$.  In this case ``just so''
is relevant for solar neutrinos and for atmospheric
neutrinos it is a general 3-flavor mixing that is operative.
 The range of mixing angles allowed has been determined by
Acker et al.\cite{Acker5}  A possible allowed matrix for example is:
\begin{equation}
U \ = \ \left(
\begin{array}{rrl}
0.64    &    0.48     &   0.6 \\
-0.76   &    0.32     &   0.56 \\
0.08    &    -0.81    &   0.57
\end{array} \right)
\end{equation}
There can be no see-saw mechanism since
there is near degeneracy.  It is even more interesting if
the degenerate mass $m$ is of order of a few $(\sim 3)$ eV.
Then the effective mass for HDM is $3m \sim 10$eV and
$\nu_e$-mass is tantalizing close to the
current $\beta-$decay mass limits \cite{Particle}.
Furthermore, in neutrinoless
double beta decay, the effective neutrino mass (assuming
Majorana neutrinos) is
\begin{equation}
\begin{array}{ccc}
<m_\nu^{eff}> & = & \ \sum_{i} \mid U_{ei} \mid^2 \eta_i m_i
 \\
              & =  &   m \Sigma \mid U_{ei}\mid^2 \eta_i
\end{array}
\end{equation}
where $\eta_i = \pm 1$ are the CP eigenvalues for $\nu_i$
and $m \sim 3$eV.  For example if $\- \eta_1 = \eta_2 =
\eta_3 = +1$ then $<m_\nu^{eff}>
\sim 0.2m \sim 0.7eV$ for the matrix above.  In any case, in
general $<m_\nu^{eff}>$ may be no smaller than a fraction of
1eV and within reach in next generation of double beta decay
experiments\cite{Raghavan1}.  Search for interesting models which yield such
interesting near degeneray, mixing patterns and mass ranges
is under way.  It has been long known that exact degenaracy
is easy to obtain\cite{Babu} by imposing symmetries; the trick is
to find the correct breaking pattern.

(iv) Another possibility which has been discussed in the
literature is the mixing of one sterile with the three
flavor neutrinos\cite{Caldwell}.  The viable scenarios are: (a) $\nu_e -
\nu_s$, MSW with small mixing for solar; and $\nu_\mu -
\nu_\tau (\delta m^2 \sim 10^{-2} eV^2)$ for atmospheric with
$m_{\nu_{\mu}} \sim m_{\nu_{\tau}} \sim 5eV$ to give
effective HDM mass of 10eV; (b) $\nu_e - \nu_\mu$, MSW for
solar; $\nu_\mu - \nu_\tau$ for atmospheric, all with masses less than
0.1 eV as discussed earlier; supplemented by $\nu_s$ of mass
of $0(10eV)$ for HDM.

(v) It should be mentioned for
completeness that the flavor violation by gravity has the
amusing feature that the same parameter range can account
for solar as well as atmospheric anomalies at the same time.

{\bf Conclusion:}

We obviously need more data!  Future solar neutrino
detectors will measure the $^8B$ spectrum and the $NC/CC$
which will be acid tests of neutrino oscillation
hypothesis; measurement of $^7Be$ will also be crucial and
the large rates will reduce statistical errors.  For the
atmospheric neutrinos; the most important milestones are:
the KEK beam test and further results from Soudan II.  Long
Baseline and reactor experiments can also test neutrino oscillation
hypothesis\cite{Reay}.  The BNL Proposal 889 is very elegant and
impressive and can test $\nu_\mu-\nu_\tau$ as well as
$\nu_\mu-\nu_e$ hypothesis.  Experiments such as CHORUS,
NOMAD and P803 will determine whether $\delta m^2_{\mu \tau}$ is in the
range 100 to 1000 $eV^2$ with even very small mixing.  The
next five years should bring many new exciting results.

{\bf Acknowledgements}

I am grateful to Andy Acker, Gary McGrath, Anjan Joshipura,
John Learned, Al Mann, Kenzo Nakamura, Jim
Pantaleone, Raju Raghavan and Atsuto Suzuki for
many enjoyable discussions, and to the organisers for
their magnificent hospitality.  This work is supported in part by
US DOE under contract DE-AM03-76SF-00325 and the University of
Hawaii.

\end{document}